# Using JSON-LD to Compose Different IoT and Cloud Services

Darko Andročec
Faculty of Organization and Informatics
University of Zagreb
Pavlinska 2, 42000 Varaždin, Croatia
dandrocec@foi.hr

*Abstract*— Internet of things and cloud computing are in the widespread use today, and often work together to accomplish complex business task and use cases. This paper propose the framework and its practical implementation to compose different things as services and cloud services. The ontology based approach and JSON-LD was used to semantically annotate both types of services, and enable the mechanism to semi-automatically compose these services. The use case and proof-of-concept application that use the proposed theoretical approach is also described in this work.

## I. INTRODUCTION

There are many business use cases where interoperation among Internet of things devices and cloud services can bring benefits to organizations. The main aim of this paper is description of method and tool for IoT and cloud interoperability. For composition of thing as a service and cloud services we have chosen AI planning method, more specifically Hierarchical Task Network (HTN planning) as the AI planning technique. Integration with Global Sensor Network (GSN) software and JSON-LD is also done in this document.

The Representational State Transfer (REST) architectural style is considered a reference paradigm in the literature to bring smart things into the Web [1]. JSON-LD [2] is a simple method to add semantics to existing JSON documents. Semantic data is serialized in a way that is often indistinguishable from traditional JSON [2]. In our framework, we use AI planning method [3], our developed open IoT ontology (https://github.com/dandrocec/IoTOntology/blob/master/ThingAsAServiceOntology.owl), and custom Java code to enable composition of IoT services. On the lower layer, things will be annotated using JSON-LD and connected to the GSN middleware [4] if they use sensors, or directly to develop Java web services, if they use actuators.

This paper proceeds as follows. First, in Section 2, the related work is listed. In Section 3 it is described how to compose different things and cloud services. Next section presents sample JSON-LD files used in our implementation. The conclusions are provided in the final section.

## II. RELATED WORK

Hur et al. [5] tackled interoperability issues between physical objects and web-centric IoT platforms. Users have to learn platform-specific data structures, message formats, and provider's APIs to deploy heterogeneous things to different IoT platforms. They presented a semantic approach to generating a service description and deploying a set of heterogeneous things to different platforms automatically. Hur et al. [5] proposed the ontology that supports the translation of a platform-specific configuration into semantic metadata using a common knowledge scheme. In this ontology, a physical object has properties (key-value pairs of static attributes of a physical object – e.g. name, description…), capability (dynamic data provided by physical object), and server profile (a configuration how a physical object can interact with a specific platform). A capability specifies the capability name, data type, data unit, and condition of a physical thing. The server profile can describe server name, URL, and the available actions of a platform. Semantic metadata of things and IoT platforms are written in JSON-LD.

Kovatsch et al. [6] showed how RESTdesc and semantic reasoning can be applied to create web IoT mashups among resource-constrained devices and services. Due to high number of various devices and different application domains, classic approaches with device IDs or API documentation and manual composition do not scale. They proposed to split the semantic description of a device into a static description of its services and a dynamic description of its current state.

Guinard and Trifa [7] propose integration method that applies REST principles to embedded devices. Things are turned into RESTful resources, or an intermediate gateway is used if computational resources of a thing is too limited or thing does not offer a RESTful interface. Smart gateways abstract the proprietary communication protocols or APIs of things and offer their functionalities accessible via RESTful API. Trifa et al. [8] also developed a common model to describe physical objects in the Web of Things. "It defines a model and Web API for Things to be followed by anyone wanting to create a product, device, service, or application for the Web of Things" [8]. Guinard also proposes the Web of Things Architecture where he proposes four main layers (access, find, share, and compose).

Desai et al. [9] proposed an IoT gateway that allows translation between messaging protocols such as XMPP, CoAP and MQTT. Semantic gateway as service (SGS) bridges low level raw sensor information with knowledge centric application services by facilitating interoperability at messaging protocol and data modelling level.

### III. COMPOSITION OF DIFFERENT THINGS

When two or more things are semantically annotated using JSON-LD, we can semi-automatically compose the two or more semantically annotated things. Also, we can inspect their non-functional properties, and decide which composition alternative is better, if we have more than one option. For example, we want to build simple application that will visually show by LED light when the temperature in the room is too low or too high.

The sample JSHOP2 domain file (generated manually) is shown below:

```
(defdomain iot (
; BEGIN compose temperature sensor and LED actuator

(:method (composeIoTServices ?sensor ?actuator)
    ()
    ((!checkSensorActuator ?sensor ?actuator))
)

(:operator    (!checkSensorActuator    ?sensor ?actuator)
    ((SemanticWebThing ?thingId1) (hasResources ?thingId1    Sensor    ?sensor) (SemanticWebThing ?thingId2)    (hasResources    ?thingId2    Actuator ?actuator) )
    ()
    ((sensorCanConnectToActuator    ?sensor ?actuator))
)

; END compose temperature sensor and LED actuator
)
)
```

Not all IoT devices support HTTP (RESTful services and JSON) or WebSockets. We will connect these types of devices to GSN (our upgraded version with web services that also support actuators), and then annotate things using JSON-LD on the server side. GSN has its own RESTful services (see Figure 5), so we can supplement it with JSON-LD annotations. Possible output format of GSN REST services are csv, json, xml, shp, and asc (source: GSN Services API Documentation - https://github.com/LSIR/gsn/wiki/gsn-services-API-documentation), so we can choose JSON and upgrade it to our JSON-LD files that connects to our open IoT ontology. Solving a planning problem in JSHOP2 is done in three steps [10, p. 2] the domain description file is compiled into Java code, the problem descriptions are converted into Java class, and the second Java class should be executed to initiate the planning process and to inspect the planning results.

Problem description file is composed of logical atoms showing the initial state and a task list [10, p. 2]. The problem description file describes available IoT services (directly from things/IoT devices or through some IoT or sensor middleware such Global Sensor Network used in our project). The task list and the initial state are created on the fly, when the user chooses some of the available actions. The initial state (a set of logical atoms) is also created programmatically. Based on the action chosen by user (task list to be executed), SAWSDL, JSON-LD files, and the IoT ontology are parsed to generate logical atoms. A SAWSDL parser was developed in Java by using EasyWSDL open-source library and its extension EasySAWSDL. Also, and JSON-LD parser Java class was developed that uses JSON and JSON-LD libraries. The class for parsing OWL ontology was implemented by using Apache Jena library. The domain description file consists of operators, methods and axioms. The domain description file is defined manually.

After the domain and problem description files were successfully created, these definitions are forwarded to a component in the prototype which invokes JSHOP2 planner to get a plan if it exists. The domain and problem descriptions are dynamically compiled into Java code, and the resulting Java files are redeployed to Glassfish server.

Additionally, the existing IoT services can also be annotated in the same way, for example some things available at some web IoT platforms can be annotated using JSON-LD and later converted into JSHOP2 axioms. If the IoT services use SOAP protocol, then we alternatively use SAWSDL files to annotate these services.

The plan given by JSHOP2 is parsed to retrieve adequate web services from SAWSDL files or JSON-LD that need to be executed. Apache CXF framework [11] or similar web service framework that supports SOAP and REST services can be used to dynamically invoke web service. This framework enables a dynamic creation of web service clients, and invokes web services with their inputs.

For composition of IoT and cloud services, we will use the same mechanism as a composition of IoT services. Cloud services are annotated by using SAWSDL and can be parsed in the same way, so we can use AI planning to semi-automatically compose IoT and cloud services. Cloud services are annotated using cloud OWL ontology, e.g. platform as a service ontology described in [12]. For AI planning process, a JSHOP2 planner was used again. The main task and difference is that we need to define new JSHOP2's domain description and problem description files to include both IoT and cloud services. The simplest example could include storing sensors' data from things to a cloud storage using available cloud providers' APIs (application programming interfaces), e.g. from specific sensor connected to Arduino Yun to Microsoft Azure storage.

## IV. SAMPLE JSON-LD FILES

Here is the definition of the sample JSON-LD file to annotate Arduino Yun with temperature sensor with terms defined in our IoT ontology (https://github.com/dandrocec/IoTOntology/blob/master/ThingAsAServiceOntology.owl).

```
{
  "@context":
    {
      "rdf":      "http://www.w3.org/1999/02/22-rdf-syntax-ns#",
      "rdfs":     "http://www.w3.org/2000/01/rdf-schema#",
      "owl": "http://www.w3.org/2002/07/owl#",
      "myont": "http://iot.foi.hr/ontologies/ThingAsAServiceOntology.owl#"
    },
  "@type": "myont.SemanticWebThing",
  "thingId": {"value": "2341"},
  "thingName": {"value": "Arduino Yun"},
  "thingDescription": {"value": "Arduino Yun with temperature sensor"},

  "myont.hasResources": {
        "@type": "myont.PhysicalObject",
        "poName": "DS18B20",
        "poDescription": "Waterproof DS18B20 Digital temperature sensor",
        "hasValues" : {
            "@type": "Output",
            "outputName" : "temperature",
            "outputDescription": "temperature in Celsius degree",
            "outputUnit": "celsius"
        }
   },
  "myont.supportsProtocols": {
      "@type": "myont.IoTProtocols",
      "proName": "WiFi",
      "proDescription" : "WiFi"
    },
  "myont.supportsProtocols": {
      "@type": "myont.IoTProtocols",
      "proName": "Ethernet",
      "proDescription" : "Ethernet"
    },
  "myont.supportsProtocols": {
      "@type": "myont.IoTProtocols",
      "proName": "SerialUSB",
      "proDescription" : "Serial communication via USB"
    },
  "myont.hasSecurityProblems":{
      "@type": "myont.IoTSecurityProblems ",
      "secName": "NetworkServicesVulnerabletoDenailOfService",
      "secDescription" : "Network services vulnerable to DoS attack on SSH, DNS, and firewall service"
    },

  "myont.hasSecurityProblems":{
      "@type": "myont.IoTSecurityProblems ",
      "secName": "InsecureSoftwareFirmare",
      "secDescription" : "Kernel version shipped with the device is outdated"
    }
}
```

The second example is a JSON-LD file to annotate littleBits cloudBit with LEDs (simple actuator) with terms defined in our IoT ontology:

```
{
  "@context":
    {
      "rdf":      "http://www.w3.org/1999/02/22-rdf-syntax-ns#",
      "rdfs":     "http://www.w3.org/2000/01/rdf-schema#",
      "owl": "http://www.w3.org/2002/07/owl#",
      "myont": "http://iot.foi.hr/ontologies/ThingAsAServiceOntology.owl#"
    },
  "@type": "myont.SemanticWebThing",
  "thingId": {"value": "2341"},
  "thingName": {"value": "littleBits cloud bit"},
  "thingDescription": {"value": "littleBits with LED"},

  "myont.hasResources": {
        "@type": "myont.Actuator",
        "poName": "02 long LED",
        "poDescription": "long white LED",
        "hasValues" : {
            "@type": "Input",
            "inputName" : "input voltage",
            "inputDescription": "input voltage in percentage",
            "inputUnit": "percent"
        }
   },
  "myont.supportsProtocols": {
      "@type": "myont.IoTProtocols",
      "proName": "WiFi",
      "proDescription" : "WiFi"
    },
  "myont.hasSecurityProblems":{
      "@type": "myont.IoTSecurityProblems",
      "secName": "NetworkServicesVulnerabletoDenailOfService",
      "secDescription" : "Network services vulnerable to DoS attack on SSH, DNS, and firewall service"
    },

  "myont.hasSecurityProblems":{
      "@type": "myont.IoTSecurityProblems",
      "secName": "InsecureSoftwareFirmare",
      "secDescription" : "Kernel version shipped with the device is outdated"
    }
}
```

Below is sample of generated JSHOP2 problem file after parsing the JSON-LD definition of the Arduino Yun with temperature sensor and littleBits cloudBit with LED actuator described in the previous section. For this purpose, we have used the following two third-party Java libraries: com.eclipsesource.minimal-json and jsonld for Java:

```
(defproblem problem iot
 (
```

```
    (SemanticWebThing SemanticWebThing_2341)
    (thingName SemanticWebThing_2341 Arduino_Yun)
    (thingDescription          SemanticWebThing_2341
Arduino_Yun_with_temperature_sensor)
    (hasResources    SemanticWebThing_2341    Sensor
DS18B20)
    (Description        Sensor        DS18B20
Waterproof_DS18B20_Digital_temperature_sensor)
    (OutputName Sensor DS18B20 temperature)
    (OutputDescription         Sensor         DS18B20
temperature_in_celsius_degree)
    (OutputUnit Sensor DS18B20 celsius)
    (supportsProtocol SemanticWebThing_2341 WiFi )
    (supportsProtocol SemanticWebThing_2341 Ethernet
)
    (supportsProtocol SemanticWebThing_2341 SerialUSB
)
    (hasSecurityProblem          SemanticWebThing_2341
NetworkServicesVulnerabletoDenailOfService )
    (hasSecurityProblem          SemanticWebThing_2341
InsecureSoftwareFirmare )
    (SemanticWebThing SemanticWebThing_1234)
    (thingName              SemanticWebThing_1234
littleBits_cloud_bit)
    (thingDescription       SemanticWebThing_1234
littleBits_with_LED)
    (hasResources    SemanticWebThing_1234    Actuator
long_LED)
    (Description Actuator long_LED long_white_LED)
    (supportsProtocol SemanticWebThing_1234 WiFi )
    (hasSecurityProblem          SemanticWebThing_1234
NetworkServicesVulnerabletoDenailOfService )
    (hasSecurityProblem          SemanticWebThing_1234
InsecureSoftwareFirmare )
  )

   ((composeIoTServices DS18B20 long_LED))

)
```

## V. CONCLUSION

This paper shows the implementation of the approach (thing as a service interoperability framework [13]) and extends it to cloud interoperability too. For composition of things as a service and cloud services we have chosen AI planning method, more specifically HTN planning tool JSHOP2. In this paper, we have described our prototype developed in Java and Netbeans IDE that enables to compose different services, including thing as services and cloud services. Things are semantically annotated using JSON-LD and concepts from our IoT ontology. These annotations are programmatically converted into AI planning files that can be used to check whether the specific composition is possible. Cloud services are annotated using SAWSDL, and programmatically converted also to JSHOP2's AI planning files, so we can use the same method to compose things and cloud services (mostly associated with cloud storage functionalities).

ACKNOWLEDGMENTS

This paper has been fully supported by the Croatian Science Foundation under the project IP-2014-09-3877.